\documentclass[preprint,review,12pt]{elsarticle}

\usepackage{amssymb}
\usepackage{tabularx}
\usepackage{booktabs}
\usepackage{setspace}
\usepackage{array, multirow}

\journal{Pattern Recognition}

\begin{document}
\begin{frontmatter}



\title{Audio, Speech, Language, \& Signal Processing for COVID-19: A Comprehensive Overview}

\author[label1,label2]{Gauri Deshpande}
\author[label1,label3]{Bj\"orn W.\ Schuller}
\address[label1]{Chair of Embedded Intelligence for Health Care and Wellbeing, University of Augsburg, Germany}
\address[label2]{TCS Research Pune, India}
\address[label3]{GLAM -- Group on Language, Audio, \& Music, Imperial College London, UK}


\begin{abstract}
The Coronavirus (COVID-19) pandemic has been the research focus world-wide in the year 2020. Several efforts, from collection of  COVID-19 patients' data to screening them for the virus's detection are taken with rigour. A major portion of COVID-19 symptoms are related to the functioning of the respiratory system, which in-turn critically influences the human speech production system. This drives the research focus towards identifying the markers of COVID-19 in speech and other human generated audio signals. In this paper, we give an overview of the speech and other audio signal, language and general signal processing-based work done using 'Artificial Intelligence' techniques to screen, diagnose, monitor, and spread the awareness about COVID-19. We also briefly describe the research related to detect according COVID-19 symptoms carried out so far. We aspire that this collective information will be useful in developing automated systems, which can help in the context of COVID-19 using non-obtrusive and easy to use modalities such as audio, speech, and language.

\end{abstract}


\begin{highlights}
\item Past research on audio, speech, language, and signal processing for COVID-19 related health problems.
\item COVID-19 screening, diagnosing, monitoring, and post-care using speech processing.
\end{highlights}


\end{frontmatter}


\section{Introduction}
More than 38 million confirmed cases of coronavirus-induced COVID-19 infected individuals are detected in more than 200 countries across the world at the time of writing of this overview. The COVID-19 pandemic  had a wide spectrum of effects on the population, ranging from no symptoms to life-threatening medical conditions. As per the world health organisation (WHO)\footnote{www.who.int}, the most common symptoms of COVID-19 are fever, dry cough, and fatigue, and the symptoms of a severe COVID-19 condition are mainly shortness of breath, loss of appetite, confusion, persistent pain or pressure in the chest, and temperature above 38 degrees Celsius. The heavy droplets generated when the infectious individual sneezes or coughs transmits the virus causing COVID-19. Even breathing and talking to someone in the close proximity of a COVID-19 infected individual can transmit the disease. With the cognizance of these transmitting factors and symptoms, every individual along with the health care professionals is required to take steps to cease the spread. 

It is vital to have an easy to use tool for screening, diagnosing, and monitoring the virus and its proliferation. An automated approach to detect and monitor the presence of COVID-19 or its symptoms could be developed using Artificial Intelligence (AI) based techniques. Although, AI techniques are still in the process of reaching a matured stage, they can be still be used for early detection of the symptoms, especially in the form of a self-care tool in reducing the spread, taking early care, and hence avoiding the severe conditions to propagate. 
As reviewed by the authors of \cite{schuller2020covid}, AI based approaches using speech and other audio modalities have several opportunities in this space. Also, the authors of \cite{deshpande2020overview} have been among the very first to identify and collate the useful AI based techniques and the efforts taken for COVID-19, right when the pandemic spread was at its peak. 

\begin{figure}[t!]
  \centering
  \includegraphics[width=0.9\linewidth]{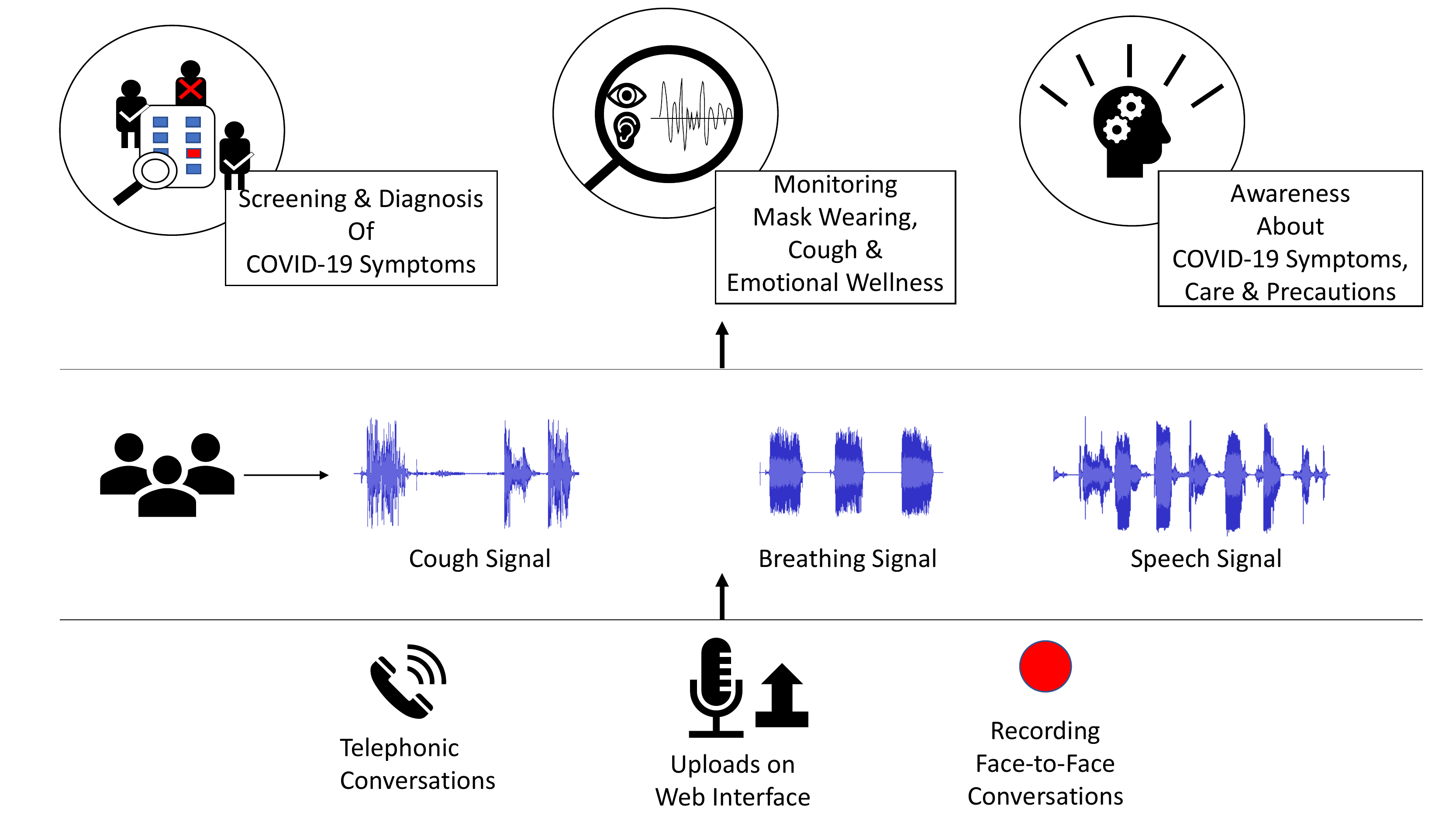}
  \caption{Capturing and processing speech and audio signals for COVID-19 applications}
  \label{fig:speechApp}
\end{figure}

As per the WHO department of mental health and substance abuse, the current scenario of COVID-19 is susceptible to elevate the rates of stress or anxiety among individuals. Especially, lock-down and social distancing, quarantine and its after effects in the society might have adverse effects on the levels of loneliness, depression, self-harm, or suicidal behaviour. A special attention is needed for the health care providers, having to face the trauma directly and spending long working hours in such scenarios. Both physical and mental health needs have increased and require AI to provide faster and easy to access solutions. Not only does identification and monitoring need digital assistance, but also the post trauma phase would need it.

As depicted in Figure \ref{fig:speechApp}, we are discussing about capturing, pre-processing and applying speech and other human audio data for screening, diagnosing, monitoring and spreading the awareness of COVID-19 in this paper. 
Tables \ref{tab:prevWork}, \ref{tab:prevWork_2}, \ref{tab:prevWork_3}, \ref{tab:prevWork_4}, and \ref{tab:prevWork_5} explain past speech and audio related work done to provide solutions for COVID-19 related health problems such as cough sound, asthmatic sound, obstructive sleep apnea, breathing rate, and stress detection from speech signals. The references given in these tables can be used readily by those who wants to develop and provide immediate solutions in this space. These tables also mention the machine learning and deep learning technologies used on the mentioned data-sets to provide mentioned accuracy. For each speech or audio application, there is a vast space in the literature. We have selected only a 
handful of them, considering their relevance in the COVID-19 situation. 
Not everything that can be developed, can be used in this scenario considering other factors such as social distancing and personal and environmental hygiene. Hence, the developments are required to be driven by guidance from the clinicians and health care providers. 

The rest of the paper is organised as follows. We start with Section \ref{sec:Screen_Diagnose} to explain the recent advances towards providing COVID-19 screening and diagnosing tools using speech and audio signals. Similarly, in Section \ref{sec:monitoring} we talk about the recent development towards the applications of speech and audio processing for monitoring the spread of the COVID-19 virus. 
In Section \ref{sec:awareness}, we describe the use of speech based technologies in the growing awareness about the pandemic so that the accurate knowledge about the disease, factors that control its expansion and preliminary care that one has to take is known to most of the individuals. We talk about the gaps where further research and development is required in Section \ref{sec:nextsteps}, and we conclude our discussion in Section \ref{sec:conclusion}.

\begin{table*}[p]
  \caption{Past speech and audio analysis related to COVID-19 health problems: Cough sound detection}
  \label{tab:prevWork}
  \centering
  \begin{tabularx}{\linewidth}{| p{1.5cm} | p{5cm} | p{6cm} |}
    \toprule
    \textbf{Ref.} & \textbf{Features} & \textbf{Dataset}\\
    \midrule
    \cite{di2017automated} & 12 Mel Frequency Cepstral Coefficients (MFCCs) & Audio data of 7 Chronic Obstructive Pulmonary Disease (COPD) patients for 90 days\\ 
    \cline{2-3}
    \textbf{Result} & \multicolumn{2}{|p{11cm}|}{Area under the curve (AUC) of around 0.916 using XGBoost.}\\
    \hline
    \cite{you2017novel} & Non-negative Matrix Factorisation (NMF) with Parameterisation using Gaussian distribution & Cough samples from 9 patients\\
    \cline{2-3}
    \textbf{Result} & \multicolumn{2}{|p{11cm}|}{1\,\% increment in accuracy, recall and f1 score
    with NMF than traditional feature sets such as MFCC, and Gammatone Frequency Cepstral Coefficients (GFCC).}\\
    \hline
    \cite{miranda2019comparative} & Short Time Fourier Transform (STFT), MFCC, Mel-filter Bank (MFB) using Deep Neural Network (DNN), convolutional Neural Network (CNN), and Long-Short Term Memory (LSTM) & Google audio set extracts from 1.8 million Youtube videos and the Freesound audio database \cite{van2016vu} containing cough signals along with other sounds such as speech, sneeze, throat clearing, and home sounds\\ 
    \cline{2-3}
    \textbf{Result} & \multicolumn{2}{|p{11cm}|}{MFB followed by STFT gives highest performance across DNN, CNN, and LSTM.}\\
    \hline
    \cite{monge2018robust} & Hu moments and k-Nearest Neighbour (k-NN) classifier & Used open source audio signals \& sounds datasets 
    and created two using data collection experiments: a) Cough sounds in noisy environments and b) 13 patients' audio recordings including cough, speech, and forced respiration.\\ 
    \cline{2-3}
    \textbf{Result} & \multicolumn{2}{|p{11cm}|}{88.51\,\% sensitivity and 99.7\,\% specificity in a variety of noise conditions.}\\
    \hline
    \cite{kvapilova2019continuous} & Spectrograms and CNN architecture & 1\,500 cough samples of 20 different subjects from 41 YouTube videos and 5 cough samples from the SoundSnap website.\\
    \cline{2-3}
    \textbf{Result} & \multicolumn{2}{|p{11cm}|}{92\,\% sensitivity at 99\,\% specificity.}\\
    
    \bottomrule
    \end{tabularx}
\end{table*}

\section{Screening and Diagnosing for COVID-19}
\label{sec:Screen_Diagnose}
In the clinical test for diagnosing COVID-19 infection, the anterior nasal swabs sample is collected as suggested in \cite{hanson2020infectious}. These tests are performed by the healthcare providers belonging to local or state healthcare departments. As the medicinal drug or vaccine for the treatment of COVID-19 is still not available, it is imperative to detect the infected individuals and physically separate them from the healthy community to stop its wide spread. Hence, additionally, as a part of any personal informatics system, the availability of self-screening or self-diagnosing methodologies can aid in detecting and self-isolating at an early stage itself.

There exists a fine line between screening and diagnosing, where screening gives an early indication of the presence of a disease and diagnosing confirms the presence/absence of disease. Screening is probabilistic, whereas diagnosis is binary in nature. To scale-up the detection of COVID-19 virus, we discuss different algorithms/applications using audio processing for screening and diagnosis of COVID-19 in this section.

\subsection{Cough Detection}
\label{subsec:cough}

\begin{figure}[t!]
  \centering
  \includegraphics[trim= 6cm 2.5cm 6cm 3cm, width=0.9\linewidth]{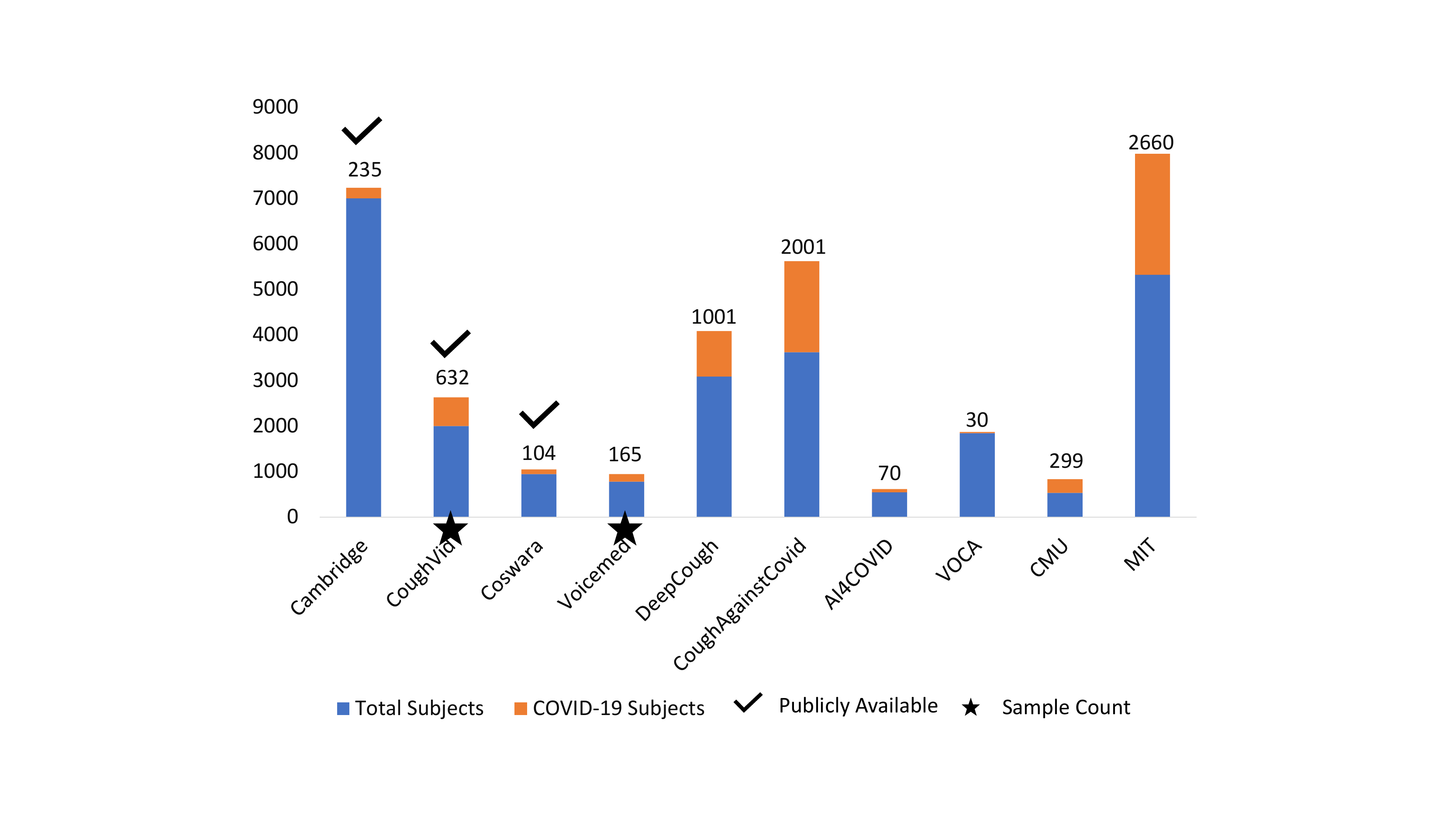}
  \caption{The Groups (mentioned on the x-axis) collected data from (the numbers mentioned on the y-axis) healthy and COVID-19 subjects. Only Coughvid and VoiceMed have reported number of samples; all others have reported number of subjects. The datasets from: Cambridge, Coughvid, and Coswara are publicly available.}
  \label{fig:coughsamples}
\end{figure}

Cough detection is about identifying the cough sound and differentiate it from other similar sounds such as speech and laughter and also, to identify COVID-19 specific cough. As a first step, it requires cough and speech samples of the same subject followed by collecting the COVID-19 and non-COVID-19 cough sound samples so as to develop an AI model that can differentiate between them on its own. Figure \ref{fig:coughsamples} shows the number of healthy and COVID-19 positive subjects or samples data collected by different groups.

Cough audio samples can be collected using a simple smartphone microphone. 
The Cambridge University\footnote{https://www.covid-19-sounds.org/en} 
provided a web based platform and an android application for the general population to upload their cough sounds along with some additional information such as their age, gender, brief medical history, location, symptoms, and if the participating subjects had been tested positive.
This data collection platform ask the participants to also read a sentence so that their speech can also be recorded. The samples comprise of 3 coughs, 5 breaths, and voice while reading via a periodic survey which captured this data after every 2 days. 
However, in the work presented in \cite{brown2020exploring}, only cough and breathing sounds are considered. 
As explained in \cite{brown2020exploring}, the crowdsourced data collected comes from more than 10 different countries and comprises of samples from more than 7\,000 subjects with more than 200 COVID-19 positive subjects. 
With manual examination of each sample, 141 cough and breathing samples of COVID-19 positive tested users and 298 samples from non-COVID-19 users (those having no medical history, no symptoms, and belonging to non-prevalent countries at that time) are used for building a binary classification model to distinguish between COVID-19 and non-COVID-19 users.
Similary, 54 ''COVID-19 with cough'' samples are distinguished from 32 ''non-COVID-19 cough'' samples, and from 20 non-COVID-19 asthmatic cough samples. 
Further, while analysing this data, the hand crafted features such as acoustic tempo, period, Root Mean Square Energy (RMSE), spectral centroid, roll-off energy, Zero Crossing Rate (ZCR), MFCC, and its derivatives are extracted. Along with the handcrafted features, the authors of \cite{brown2020exploring} used another approach as well, in which
transfer learning using the VGGish model is developed using videos from YouTube. The authors achieved an AUC of 0.8 for distinguishing COVID-19 subjects from non-COVIDs using logistic regression and again a AUC of 0.8 for distinguishing COVID-19 cough from non-COVID-19 and asthmatic cough using a support vector machine. The authors found handcrafted features along with VGGish based features to give the best performance. Together, cough and breathing signals perform the best in classifying COVID-19 users from non-COVID-19 users. However, breathing signals alone are better suited for classifying COVID-19 positive users from non-COVID-19 users having asthmatic cough. With data augmentation, the authors could improve the results by 7-8\,\%.

\begin{table*}[p]
  \caption{Past speech and audio analysis related to COVID-19 health problems: Cough sound detection (continued)}
  \label{tab:prevWork_2}
  \centering
  \begin{tabularx}{\linewidth}{| p{1.5cm} | p{5cm} | p{6cm} |}
    \toprule
    \textbf{Ref.} & \textbf{Features} & \textbf{Dataset}\\
    \midrule
    
    \cite{vhaduri2020nocturnal} & MFCC with RandomForest Classifier (RFC). & Forced cough and snore sounds from 26 healthy subjects superimposed with the AC noise.\\
    \cline{2-3}
    \textbf{Result} & \multicolumn{2}{|p{11cm}|}{An average accuracy of 0.96\,±\,0.08, F1 score of 0.96\,±\,0.08, and AUC-Receiver Operating characteristics (ROC) of 0.98\,±\,0.04 in classifying snore and cough sounds}\\
    \hline
    \cite{amoh2016deep} & Visual analysis using CNN and sequential analysis using Recurrent Neural Network (RNN) & Audio signals captured using a piezo sensor of 14 healthy subjects comprising of cough, speech, and other sounds\\
    \cline{2-3}
    \textbf{Result} & \multicolumn{2}{|p{11cm}|}{CNN Specificity: 92.2\,\% \& RNN sensitivity: 87.7\,\%.}\\
    \hline
    \cite{barata2019towards} & CNN with device-agnostic bagging ensemble method & Cough, speech, laughter, throat clearing, and forced expiration sounds of 43 healthy participants.\\
    \cline{2-3}
    \textbf{Result} & \multicolumn{2}{|p{11cm}|}{Mean value range 85.9\,\% to 90.9\,\% over 5 different recording devices.}\\
    \hline
    \cite{larson2011accurate} & Principal Component Analysis (PCA) on audio spectrograms, Fast Fourier Transfor (FFT) coefficients, and RFC & Acted cough from 17 patients having a cough due to common cold (n=8), asthma (n=3), allergies (n=1), and chronic cough (n=5)\\
    \cline{2-3}
    \textbf{Result} & \multicolumn{2}{|p{11cm}|}{True positive rate: 92\,\%, false positive rate: 0.5\,\%.}\\
    \hline
    \cite{pramono2019automatic} & Three spectral band features along with a logistic regression classifier & 43 real-world environment recordings from the video sources of \cite{pramono2016cough}, with sounds of COPD, pertussis, croup, common cold, bronchitis, bronchiolitis \& asthma.\\
    \cline{2-3}
    \textbf{Result} & \multicolumn{2}{|p{11cm}|}{Sensitivity: 90.31\,\%, specificity: 98.14\,\%, F1-score: 88.70\,\%.}\\
    \hline
    \cite{hoyos2018efficient} & Local image (Hu) moments over audio spectrograms & 13 subjects samples with 3 respiratory conditions for 1 day having low, medium, and high noisy background.\\ 
    \cline{2-3}
    \textbf{Result} & \multicolumn{2}{|p{11cm}|}{Smartphone based cough detector having 88.94\,\% sensitivity and 98.64\,\% specificity in noisy environments with optimal battery consumption.}\\
    
    \bottomrule
    \end{tabularx}
\end{table*}

Coughvid\footnote{https://coughvid.epfl.ch} is another such app from EPFL (Ecole Polytechnique Fédérale de Lausanne) to detect COVID-19 cough from other cough categories such as normal cold and seasonal allergies. Till date, this dataset \cite{orlandic2020coughvid} has more than 20\,000 cough samples collected, where all the samples are passed through an open-source cough detection machine learning model to identify the cough segments. More than 2\,000 samples are labelled by 3 expert pulmonologists for the respiratory conditions along with the COVID-19 status. The dataset has 632 COVID-19 labels given by expert 1 alone, however, there exists no agreement on the COVID-19 diagnosis (Fleiss' Kappa score 0.00).
This dataset is made publicly available along with a machine learning model for identifying cough from other sounds.

Another such effort called, 'Breath for Science' \footnote{https://www.breatheforscience.com} -- a team of scientists from NYU --, have developed a web based portal to register the participants where they can enter similar details along with a phone number. On pressing a 'call me' button, the participants receive a callback where they have to cough three times after the beep. They have mentioned that this is available for only US citizens as of now. 

As explained in \cite{sharma2020coswara}, another corpus called ''Coswara'' of 941 subjects and 9 different sounds is formed using a web interface developed by IISC Bangalore India. The nine sounds include, shallow and deep cough, shallow and deep breathing, the vowels /ey/, /i/, and /u/, and one to twenty digit counting in normal and fast speaking rate. The metadata collected from the participants includes age, gender, location, current health status (healthy / exposed / cured / infected) and the presence of co-morbidity. Here, along with the subjects labelling their own data, one more annotation is provided by the listener to assign it to one of the 9 categories. The dataset comprises of audio samples from 104 unhealthy users. After curating, the dataset is publicly available at Github\footnote{https://github.com/iiscleap/Coswara-Data}. The authors achived an accuracy of 66.74\,\% in classifying the samples into 9 categories using acoustic features such as spectral contrast, 13 MFCCs, spectral roll-off, spectral centroid, mean square energy, polynomial fit to the spectrum, ZCR, spectral bandwidth, and spectral flatness.
A subset of Coswara along with data gathered using the Stanford University led Virufy mobile app\footnote{Virufy, http://archive.is/hbrfE} and Google's AudioSet \cite{gemmeke2017audio} is used to train a MFCC-CNN model for identifying COVID-19 cough \cite{dunne2020high}. The authors have used a total of 16 COVID-19 cough samples and achieved an accuracy of 100\,\% in the classification task, however, the dataset considered 
too small to draw conclusions for generalisation ability in real-life settings.

VoiceMed\footnote{https://voicemed-791a3.firebaseapp.com} is another android and web application which captures crowd-sourced speech and cough sounds and returns the COVID-19 infection status on the fly. The different stages in this cloud-based pre-trained CNN based system comprises of pre-processing the collected signal, using a cough detector to identify if it is a cough signal and then a COVID-19 cough detector to further detect if the audio signal is a COVID-19 cough. As explained in \footnote{https://health-sounds.cl.cam.ac.uk/workshop20/Thayabaran\_Kathiresan.mp4}, the authors used 900 cough and 2\,000 non-cough audio samples for building the cough detector. Similarly, the authors have used 165 COVID-19 and 613 non-COVID-19 samples for building the COVID-19 cough detector. The accuracy of the cough classifier is reported to be 83.7\,\% and the accuracy of COVID-19 classifier is reported as 89.69\,\% using spectrograms. The authors mention that identifying older COVID-19 patients and COVID-19 patients with respiratory disorders is a further complex problem. 

However, as opposed to the self annotated data, the authors of \cite{andreu2020novel} collected a dataset of 3\,087 clinically validated samples, in which, 1\,001 positive and 2\,086 negative COVID-19 samples are present. The feature set used for cough classification are, MFCC, ZCR, roll-off frequency, and spectral centroid. The authors have described a deep convolutional architecture termed as 'DeepCough', which is further deployed as a web based pre-screening tool called 'CoughDetect' \footnote{https://coughdetect.com}.

Another web interface 'CoughAgainstCovid' \footnote{https://www.coughagainstcovid.org} for COVID-19 cough sample collection is an initiative by the Wadhwani AI group in collaboration with the  Stanford University\footnote{https://www.stanford.edu}. As explained in \cite{bagad2020cough}, the authors have collected the cough sounds forcefully produced by 3\,621 individuals using a smartphone microphone in the setups established at testing facilities and isolation wards across India. This dataset contains data from 2\,001 COVID-19 positive subjects, where these subjects' Reverse Transcription–Polymerase Chain Reaction (RT-PCR) tests are also conducted for confirmation. The authors have also developed a CNN architecture based model for classification of COVID-19 cough from Non-COVID-19 cough sound. The feature set of RMSE, tempo, and MFCCs is giving a specificity of 31\,\% for a sensitivity of 90\,\%.

Similarly, cellular call recordings of 88 subjects, with 29 positive and 59 negative COVID-19 clinically tested individuals is collected in \cite{pinkas2020sars}. The subjects were asked to provide the speech of /ah/ and /z/, counting from 50 to 80 and coughs for 14 days. The authors have compared the performance using three deep learning components, attention based transformer, a GRU-based expert classifier with aggressive regularisation and ensemble stacking. The authors report that the performance with /z/ phoneme is better when compared to /ah/ and counting. Among the deep learning techniques, transformer based experiments gave better F1 scores.

\begin{table*}[p]
  \caption{Past speech and audio analysis related to COVID-19 health problems: Asthmatic sound detection}
  \label{tab:prevWork_3}
  \centering
  \begin{tabularx}{\linewidth}{| p{1.5cm} | p{5cm} | p{6cm} |}
    \toprule
    \textbf{Ref.} & \textbf{Features} & \textbf{Dataset}\\
    \midrule
    
    \cite{yadav2020analysis} & INTERSPEECH 2013 Computational Paralinguistics Challenge baseline acoustic features \cite{schuller2013interspeech} & Speech from 47 asthmatic and 48 healthy controls\\
    \cline{2-3}
    \textbf{Result} & \multicolumn{2}{|p{11cm}|}{78\,\% Accuracy.}\\
    \hline
    \cite{sharma2018disease} & MFCC, pitch, intensity, jitter, shimmer, formants, harmonicity, fundamental frequency features with Dynamic Time Warping (DTW) & Vowel pronunciation samples of 21 asthmatic and 21 non-asthmatic individuals\\
    \cline{2-3}
    \textbf{Result} & \multicolumn{2}{|p{11cm}|}{Markers of asthmatic individuals: Lower pitch, higher standard deviation of pitch, higher degree of voice breaks, lower intensity, shimmer value greater than 3.8, higher jitter, average Harmonics to Noise Ratio (HNR) of 14.4, higher F1, and lower F2.}\\
    \hline
    \cite{lin2016automatic} & MFCCs with Gaussian Mixture Model (GMM) & Respiratory sound of 9 asthmatic and 9 non-asthmatic adults\\
    \cline{2-3}
    \textbf{Result} & \multicolumn{2}{|p{11cm}|}{Sensitivity of 0.881 and a specificity of 0.995.}\\
    \hline
    \cite{kutor2019speech} & Correlation between HNR of speech signal and Forced Expiratory Volume to Forced Vital Capacity (FEV1/FVC) ratio obtained from spirometry & Spirometry data, Vowel (/a:/, /e:/, /i:/, /o:/, /u:/, /epsilon:/, /open-o:/) pronunciation and phrase ``She sells'' of 150 asthmatic subjects, analysis done on 33 samples.\\
    \cline{2-3}
    \textbf{Result} & \multicolumn{2}{|p{11cm}|}{Highest correlation coefficient of 42.08 found between the (FEV1/FVC) ratio and HNR of vowel /epsilon:/ sound.}\\
    \hline
    \cite{nathan2019assessment} & Speech features: pause time and frequency, Prosodic features: absolute and relative jitter and shimmer & Samples from 91 with asthma and/or COPD, and 40 healthy controls.\\
    \cline{2-3}
    \textbf{Result} & \multicolumn{2}{|p{11cm}|}{68\,\% accuracy in differentiating patients from healthy individuals and 89\,\% accuracy in differentiating the subset of patients with the highest disease severity from healthy ones.}\\
    \bottomrule
  \end{tabularx}
\end{table*}

A cloud based smartphone app is described in \cite{imran2020ai4covid}, for detecting COVID-19 cough.
As a first step, the authors have used a CNN based cough detector, which identifies cough sounds from over 50 environmental sounds. The authors built this detector using the ESC-50 dataset \cite{piczak2015esc}. In the next stage, to diagnose a COVID-19 cough, they collected 96 bronchitis, 130 pertussis, 70 COVID-19, and 247 normal cough samples to train their COVID-19 cough detector model. Using MFCCs for feature representation and t-distributed stochastic neighbour embedding for dimensionality reduction, they have trained three models: a deep transfer learning-based multi-class classifier (using a CNN), a classical machine learning-based multi-class classifier (using a Support Vector Machine (SVM)), and deep transfer learning based binary classifier (again using a CNN). These three models reside in the AI4COVID engine, where a decision is made as COVID-19 positive or negative if the output of all the three models' outputs are the same; else, it says that the test is inconclusive. With this, the authors report an accuracy of more than 95\,\% in identifying cough sounds from non-cough sounds. The three engine-based models yield an accuracy of 92.64\,\%, 88\,\% and 92.85\,\% respectively for detecting a COVID-19 cough sound. The overall performance indicates that the app can detect COVID-19 infected individual with a probability of 77.3\,\%.

The COVID-19 cough data collection at Massachusetts Institute of Technology (MIT) is done using the web app \footnote{opensigma.mit.edu}, in which each subject gave 3 forced cough recordings, diagnosis details, and other demographic metadata. The authors have used MFCCs with a CNN architecture, and ResNets to build a baseline model using the collected dataset. To understand the impact on COVID-19 diagnosis by the four bio-markers: muscle degradation, vocal cords, sentiments, and lung \& respiratory tract, the authors of \cite{laguarta2020covid} have used these bio-markers as a pre-training step for the baseline model. 
This baseline model's performance is then compared with the four variants, in which it is found that the lung \& respiratory tract bio-markers have the most, and the sentiment bio-marker has the least effect on improving the baseline performance of detecting COVID-19 cough samples. The authors have reported an accuracy of 98.5\,\% in detecting COVID-19 cough, however, the data used for building the models is not clinically validated, hence they intend to work with clinically validated data next.

The winners of the 72h online hackathon ''\#CodeVsCOVID19'' which was organised in March-2020, have developed a COVID-19 cough detector using 400 cough samples \footnote{https://detectnow.org}. They used a re-trainable RandomForest classifier for the COVID-19 cough detector. Their plans are to move from crowd-sourced data to clinically validated data. 

Although the above mentioned algorithms are attaining good results on their respective test datasets, it is essential to validate these systems by using them in real time. From the hygiene perspective, it is not advised to cough on open surface. As explained in \cite{singhal2020review}, the infection is primarily transmitted through large droplets generated during coughing and sneezing by symptomatic and also by asymptomatic individuals before an onset of symptoms. These infected droplets can spread 1-2\,m, deposit on surfaces, and can remain viable for days in favourable atmospheric conditions, but can be destroyed in a minute by common disinfectants. Hence, for the collection procedures, it is important to remind the participants to cover the mouth and then only provide the cough sound samples.  Otherwise, this may result in further spreading of the disease. Also, after giving the samples, the smartphone surface should be applied with a disinfectant. 

Some initiatives have focused on identifying the symptoms instead of working on the COVID-19 positive users data.
A web based application (available only in the US)\footnote{https://www.coughmode.com}, trained a running nose cough detector, which is one of the COVID-19 symptoms, using 2\,352 cough samples from 193 subjects. They found that the data was easily separable, however, the data collected for other symptoms such as difficulty in breathing and sore throat was not forming distinguishable clusters.
A system proposed in \cite{petrellis2020covid}, comprises of a 'client' mobile and a 'supervisor' desktop app. A FFT-based analysis is used to classify the cough carrying COVID-19 symptoms such as dry and wet cough with an accuracy of around 90\,\%.
A headset based system is proposed in \cite{stojanovic2020headset}, to capture noise robust audio signals for detecting cough sounds and tracking other COVID-19 symptoms such as the respiration rate. 
The data collected from Google's audioset and ESC-50 is used by the authors of \cite{bansal2020cough}, in which they manually labelled the samples under dry-cough category as COVID-19 cough and attempted to classify these samples from others. With MFCCs as features and CNN architecture, the authors report an accuracy of 70.58\,\% in classifying dry cough from others.

\begin{figure}[t!]
  \centering
  \includegraphics[trim=4.5cm 7.5cm 4.5cm 4.5cm, width=0.7\linewidth, angle=270]{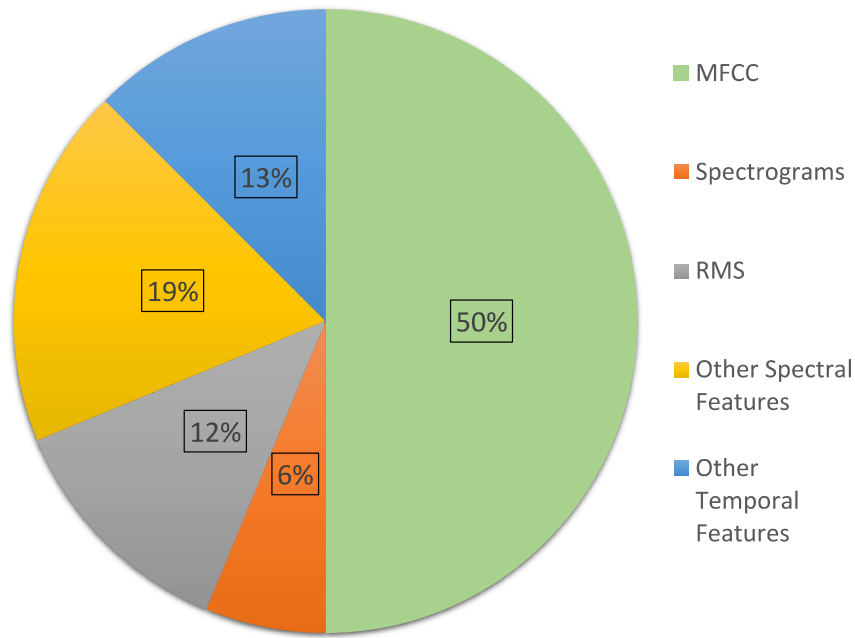}
  \caption{Acoustic features' usage for detecting COVID-19}
  \label{fig:surveyfeatures}
\end{figure}

\subsection{Speech Analysis}
\label{subsec:speech}
Considering the magnifying effects of coughing sound  
on spreading the infection in the absence of any preventive measures, capturing and analysing speech signals is a dependable alternative. A web interface to capture speech along with cough of COVID-19 patients is developed by Voca\footnote{https://voca.ai/corona-virus}. The data collected has been analysed by the author of \cite{dubnov2020signal}, where the data comprises of 30 positively diagnosed and 1\,811 healthy participants' speech and cough samples. While speaking, the candidates are asked to speak specifically 'Ahhh', 'Ehhh', 'Ohhh', 1-20 counting, a--z alphabet, and reading a segment from a story. In classifying sick individuals from the healthy ones, the author reached a maximum of 70\,\% accuracy using MFCC features and a CNN architecture.

As seen in Figure \ref{fig:surveyfeatures}, MFCCs are used in almost 50\,\% of the total efforts. However, a study  done in \cite{alsabek2020studying} with MFCCs extracted from the cough, deep breath and speech signals from 7 COVID-19 patients and 7 healthy individuals shows that MFCCs from speech are not dependable features for classifying COVID-19 and healthy individuals. Hence, it is required to understand the exact speech-based features to differentiate COVID-19 patients from healthy individuals.
In the work driven by Carnegie Mellon University (CMU), the features from models of voice production are explored to understand the presence of COVID-19 symptoms from the speech signals. These features, that the authors used for this analysis, are described in \cite{zhao2020speech}, which is based on the ADLES (Adjoint Least-Squares) method  algorithm, which extracts the features representing the oscillatory nature of the vocal fold while pronouncing phonemes. The voice production model that is referred to here is called the ``asymmetric body-cover'' model that estimates parameters such as the glottal pressure, mass, spring, and damping from the left and right vocal folds motion speed and acceleration. 
CMU's data collection happened through the web interface along with certain collaborations with a clinical setting. Through these collaborations, they could collect 530 subjects' data, all of which went through a COVID-19 test to validate; also, their recordings were manually vetted. The data comprises of signals from 299 positive and 231 negative tested subjects. Each subject provided 6 recordings, which includes alphabets, counting 1-20, extended vowel sounds, and coughs. 
From their observations, they have concluded that the results with extended vowels are better than that with cough signals. Authors of \cite{deshmukh2020interpreting} have explored the significant features for the detection of COVID-19 with the focus on voice production process. They have analysed the differential dynamics of the glottal flow waveform (GFW) during voice production with the recorded speech, as it is not possible to analyse the GFW of COVID-19 patients. They hypothesize that a greater similarity between the two indicates normal voice and the larger difference would mean the presence of anomalies. A CNN based 2-step attention model is used to detect these anomalies from the 19 subjects' extended vowels /a/, /i/, /u/ recordings, in which 9 recordings belong to COVID-19 positive subjects. The authors report the residual and the phase difference between the two GFWs as the most promising features yielding the best classification AUC of 0.9 on the extended vowels /i/+/u/.

Another web based interface for detecting COVID-19 symptoms from the voice is the  ''Spira Project'' \footnote{https://spira.ime.usp.br/coleta}. This interface asks the participants to record three phrases. In describing their initial results using MFCCs and a CNN architecture, the authors reported an accuracy of 91\,\% in detecting COVID-19 symptoms related to respiratory disorders.

The Afeka college of engineering has developed a mobile application for remote pre-diagnostic assesment of COVID-19 symptoms from the voice and speech signals captured from infected and healthy individuals. Their training dataset comprises of 70 speakers and 235 events in the training, and 18 speakers and 57 events in the test set \footnote{https://health-sounds.cl.cam.ac.uk/workshop20/Alon\_Barnea\_Vered\_Aharonson.mp4}. They achieved precision of 0.79 on the test set.

Speech recordings of TV interviews of COVID-19 positive speakers available on YouTube are collected and analysed in \cite{ritwik2020covid} for classifying COVID-19 patients from healthy individuals. The data-set is publicly available \footnote{https://github.com/shareefbabu/covid\_data\_telephone\_band} which comprises of 19 speakers with 10 of them tested COVID-19 positive. The data collected is manually segmented, after which MFB features are calculated for the speech segments. Using the ASpIRE chain model, which is a time delayed neural network trained on the Fisher English dataset described in \cite{ko2015audio}, the authors of \cite{ritwik2020covid} have extracted the posterior probability of phonemes for each frame. When concatenated, this gives a feature vector for each speech utterance. Using a SVM classifier, the authors report an accuracy of 88.6\,\% and F1-score of 92.7\,\% in classifying COVID-19 patients from healthy speakers.

The authors of \cite{pal2020pay} discuss the interpretability of their framework of COVID-19 diagnosis using embeddings for the cough features and symptoms' metadata. In this study, cough, breathing, and speech of counting from 1-10 is collected from 150 subjects, in which 100 subjects were COVID-19 positive, and 50 were tested negative during their RT-PCR test. Apart from this, the authors also collected data for bronchitis and asthma cough from online and offline sources. The authors report an improvement of 5-6\,\% in the accuracy, F1-score, sensitivity, specificity, and precision when using both the symptoms'  metadata and cough features for the classification tasks.
Similarly, the authors of \cite{belkacem2020end} have used additional information such as: airflow from spirometer, body temperature from thermal camera, heart rate from ECG, electrical activity that causes  muscle contraction around the heart and chest region, along with cough detection from a microphone to build a sensor-based system for identifying COVID-19 subjects.

\begin{table*}[ht!]
  \caption{Past speech and audio analysis related to COVID-19 health problems: Speech breathing analysis and stress detection}
  \label{tab:prevWork_4}
  \centering
  \begin{tabularx}{\linewidth}{| p{1.5cm} | p{5cm} | p{6cm} |}
    \toprule
    \textbf{Ref.} & \textbf{Features} & \textbf{Dataset}\\
    \midrule
    
    \cite{routray2019automatic} & Cepstrogram and SVM with radial basis function & Speech recordings of 16 participants of age group 21 years mean\\
    \cline{2-3}
    \textbf{Result} & \multicolumn{2}{|p{11cm}|}{89\,\% F1 score and RMSE of 4.5 breaths/min for the speech-breathing rate.}\\
    \hline
    \cite{nallanthighal2019deep} & Spectrogram with a CNN and a RNN & 20 healthy subjects' speech recorded using microphone and breathing signal using two respiratory transducer belts\\
    \cline{2-3}
    \textbf{Result} & \multicolumn{2}{|p{11cm}|}{91.2\,\% sensitivity for breath event detection and mean absolute error of 1.01 breaths per minute for breathing rate estimation (Breathing signal detection for conversational speech)}\\
    \hline
    \cite{azam2018smartphone} & Wavelet de-noising and Empirical Mode Decomposition for data pre-processing. Instantaneous frequency and envelop features, and SVM classifier. & 255 breath cycles captured using smartphone microphone\\
    \cline{2-3}
    \textbf{Result} & \multicolumn{2}{|p{11cm}|}{Accuracy of (75.21\,\%\,±2\,\%) for asthmatic inspiratory cycles and (75.5\,\%\,±3\,\% for complete respiratory sounds cycle with a diagnostic odds ratio of 20.61\,\% and 13.87\,\% respectively.}\\
    
    \hline
    \cite{partila2019human} & Low Level Descriptors and functional features extracted using openSMILE \cite{eyben2010opensmile} with k-NN, SVM, and CNN classifiers & 31 emergency call recordings of the Integrated Rescue System of the 112 emergency line from the Czech Republic\\
    \cline{2-3}
    \textbf{Result} & \multicolumn{2}{|p{11cm}|}{Accuracy of 87.9\,\% with SVM and 87.5 \% with CNN in classifying stress from neutral speech}\\
    \bottomrule
  \end{tabularx}
\end{table*}

\subsection{Breathing Analysis}
\label{breathing}
As discussed before, shortness of breath is also one of the symptoms of the virus for which smartphone apps are designed to capture breathing patterns by recording the speech signal. As described in Section \ref{subsec:cough} and \ref{subsec:speech}, multiple attempts are made to analyse respiration along with cough and speech signals. Especially, the authors of \cite{brown2020exploring} have reported that the breathing signals are better suited for classifying COVID-19 positive users from healthy users having asthma and a cough. In an another study of \cite{hassan2020covid} with a small dataset of 60 healthy and 20 COVID-19 positive subjects, the authors report better accuracy with RNN based analysis using both breathing and cough data than that of speech. The feature set used by the authors include spectral centroid, spectral roll-off, zero crossing rate, MFCCs and their derivatives.
However, as compared to cough classification, analysing breathing signals are less popular owing to the challenges in capturing noiseless breathing signals.
There have also been efforts taken in correlating speech signals with the breathing signals.
In the Breathing Sub-challenge of Interspeech 2020 ComParE \cite{schuller2020interspeech}, the authors have achieved a baseline pearson's correlation of 0.507 on a development, and 0.731 on the test dataset, respectively. They have used two piezoelectric respiratory belts for capturing breathing patterns. 

In an another effort of correlating speech signals with breathing signals, an ensemble system with fusion at both feature and decision level of two approaches is presented in \cite{markitantov2020ensembling}. One of the two approaches is a 1D-CNN based end-to-end model having two LSTM layers stacked above it. The other approach uses a pre-trained 2D-CNN ResNet18 with two Gated Recurrent Unit (GRU) layers stacked above it. The fusion that happens at feature level combines the embeddings at the last layers of the 1D-CNN and the ResNet18 to train a two-layer LSTM network. The decision level fusion combines the predictions from the 1D-CNN, ResNet18, and the feature level fusion-based approaches. The authors have reported a Pearson's correlation coefficient (r-value) of 0.763 between the speech signal and corresponding breathing values of the test set.

Further, the authors of \cite{mendoncca2020analyzing}, modified the end-to-end baseline architecture, by replacing the LSTMs by Bi-LSTM. The authors also augmented the challenge dataset, with the same dataset being modified to emulate the Voice over Internet (VoIP) conditions. With the above modifications, they achieved an r-value of 0.728 on the test dataset.

To explore attention mechanisms, the authors of \cite{macintyre2020deep} have used an end-to-end approach along with a  Convolutional RNN (CRNN) for two prediction tasks: the breathing signals captured using a respiratory belt, and the inhalation events. The authors report a maximum of 0.731 r-value in predicting the breathing pattern from the speech signal and the macro averaged F1 value of 75.47\,\% in predicting the inhalation events. The attention step is found to improve the metrics by 0.003 r-value, and .726\,\% F1 value for the two tasks, respectively.

Outside of the challenge, the authors of \cite{nallanthighal2020speech} have attempted to correlate high quality speech signals captured using an Earthworks microphone M23 at 48\,kHz with the breathing signal captured using two NeXus respiratory inductance plethysmography belts over the ribcage and abdomen to measure the changes in the cross-sectional area of the  ribcage and abdomen at a sample rate of 2\,kHz. The authors have achieved a correlation of 0.42 to the actual breathing signal and a breathing error rate of 5.6\,\% and sensitivity of 0.88 for breath event detection. 

\begin{table*}[p]
  \caption{Past speech and audio analysis related to COVID-19 health problems: Obstructive Sleep Apnea Detection (OSA)}
  \label{tab:prevWork_5}
  \centering
  \begin{tabularx}{\linewidth}{| p{1.5cm} | p{6.5cm} | p{4.5cm} |}
    \toprule
    \textbf{Ref.} & \textbf{Features} & \textbf{Dataset}\\
    \midrule
    
    \cite{simply2018obstructive} & Breathing detection: MFCCs with a single layer NN, OSA classification: MFCCs, energy, pitch, kurtosis,  and ZCR with SVM. & 90 Male subjects' speech and sleep quality measures using WatchPAT \cite{gan2017validation}\\
    \cline{2-3}
    \textbf{Result} & \multicolumn{2}{|p{11cm}|}{Cohen's kappa coefficient of 0.5 for breathing detection and 0.54 for OSA detection.}\\
    \hline
    \cite{botelho2019speech} & Formant frequencies and bandwidth (F1, F2, and F3), HNR, jitter, spectral flux, F0, MFCCs, and Linear prediction cepstral coefficients with an ensemble of SVM, Linear Discriminant Analysis, and k-NN & Speech of 25 OSA subjects and 20 controls\\
    \cline{2-3}
    \textbf{Result} & \multicolumn{2}{|p{11cm}|}{True-Positive-Rate of 88\,\%, and a True-Negative-Rate of 80\,\% verified on an in-the-wild data-set acquired from YouTube.}\\
    \hline
    \cite{espinoza2016reviewing} & i-vector techniques with Support Vector Regression & OSA suspected 426 male Spanish speakers speech\\
    \cline{2-3}
    \textbf{Result} & \multicolumn{2}{|p{11cm}|}{Poor results due to limiting factors such as high correlation of the speech signal with parameters such as age, sex, and others and over-fitting of models with lengthy feature-sets.}\\
    \hline
    \cite{mendonca2018review} & Review paper & 84 papers reviewed on OSA from 2003 to 2017\\
    \cline{2-3}
    \textbf{Result} & \multicolumn{2}{|p{11cm}|}{ECG-based analysis supersedes sound-based analysis for OSA detection, both for data availability and accuracy perspective. This can be attributed to the availability of the cleaner ECG signal as compared to noisy respiratory sounds and speech signals. Most prominently used ML techniques include SVM, k-NN, and Neural Network.}\\
    \hline
    \cite{qian2020can} & Authors have reviewed the Machine Learning (ML) techniques in locating the excitation of snore sound & 23 papers from 2015 to 2019\\
    \cline{2-3}
    \textbf{Result} & \multicolumn{2}{|p{11cm}|}{Audio features along with traditional ML techniques and state of the art deep learning techniques are reviewed. Due to absence of fundamental knowledge regarding the acoustic properties of snoring sounds and publicly available datasets, the research in this direction has not progressed much.}\\
    \bottomrule
  \end{tabularx}
\end{table*}

The feasibility of capturing and analysing breathing sounds using smartphone microphones to even understand COVID-19 symptoms as a part of a personal analytics is discussed in \cite{faezipour2020smartphone}. 
A preliminary analysis of the sound signals of respiration from 9 COVID-19 patients and 4 healthy volunteers is executed in \cite{furman2020remote} using FFT harmonics. The respiration sounds are recorded using a smartphone microphone. The rule based analysis shows distinction in the spectrum of healthy and infected individuals, however, more data from more subjects are required to strengthen the analysis. The Cambridge University app that captures breathing sounds along with cough and speech features a COVID-19 breath analyser as well. Another such app detecting anomalies from the breathing sound has been developed by the TCS Research team \cite{harinis2020}.
A MFCC based breath detector detects the breath sound. Further, MFCC along with Power Spectral Density (PSD) and spectrograms are used in an anomaly detection engine which finds an anomaly present in the breath sound. The anomaly detection engine is built using SVM. The dataset used in this project comprises of data donated by TCS internals along with that from the 
RALE repository \cite{beck2008rale} of lung sounds. The authors have demonstrated that their app can detect COVID-19 anomalies as well.

\subsection{Chat-Bots}
\label{chatbot-S}
As the count of positive COVID-19 cases are increasing every day, it brings up the need of automating the conversation a physiologist would have to understand the presence of symptoms in an individual. 
Microsoft and CDC have come up with a chatbot named ``Clara'' for initial screening of COVID-19 patients by asking them questions and capturing the responses. This uses speech recognition and speech synthesis technologies. The risk-assessment test is designed based on advice from the WHO and the Union Ministry of Health and Family Welfare India\footnote{https://www.mohfw.gov.in}. ``Siri'' from Apple is also updated to answer the variations of the general question, 
``Siri, do I have the coronavirus?'' based on WHO guidelines. If a person shows severe symptoms, then it is advised by Siri to call 911. Similarly, Amazon's Alexa is also updated with answering COVID-19 screening questions based on CDC guidelines.
Considering the trauma that the health care providers are going through, a web-based chat-bot named Symptoma developed in \cite{martin2020artificial}, is a significant step. It can differentiate 20\,000 diseases with an accuracy of more than 90\,\% and can identify COVID-19 from other diseases having similar symptoms with an accuracy of 96.32\,\%. It uses semantic analysis on the free text data entered by the user.

\begin{figure}[t!]
  \centering
  \includegraphics[width=\linewidth]{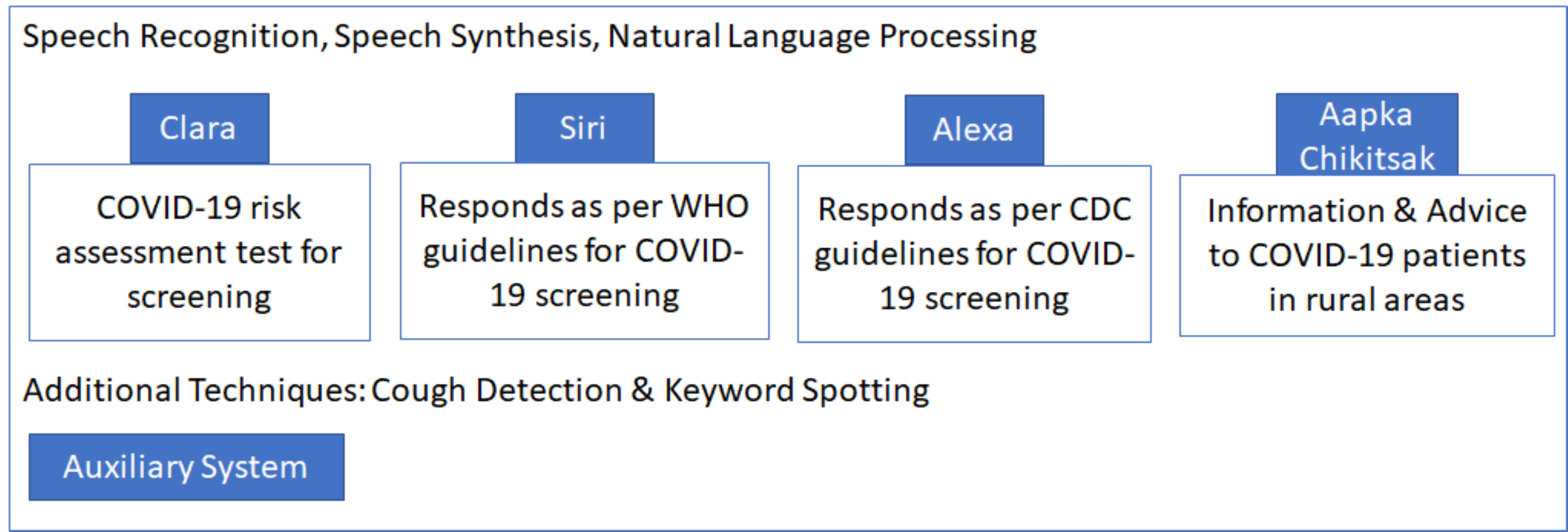}
  \caption{Chat-bots in COVID-19 context}
  \label{fig:chatbots}
\end{figure}

The robot based system for COVID-19 risk analysis mentioned in \cite{wei2020real} uses speech and audio analysis techniques such as speech recognition, keyword detection, and cough detection and classification in real-time. The authors have used a CNN architecture for detection of cough events from robot-human conversations and a classifier based on attentional similarity for detecting a COVID-19 infection. They have 1\,283 speech recording segments of the interaction of 184 respondents aged 6 to 80 years, which includes 392 segments from 64 lab confirmed COVID-19 patients. The rest of the data includes segments from respondents with long smoking history, acute bronchitis, chronic pharyngitis, and children with pertussis along with healthy candidates. 21 COVID-19 candidates are having other chronic diseases such as hypertension, diabetes, and heart related diseases. 
Before the conversation starts between the robot and the candidate, other measures such as temperature and demographic information are collected, stored and analysed. During the conversation, the robot monitors the data for cough signals, if not received, it asks the candidate to cough forcefully towards the end of the conversation. At the end, a report is generated and shared with the candidate and authorised doctors.

\subsection{Tele-Health}
The increasing count of COVID-19 infectants among healthcare providers suggest the need of telehealth systems where the required care, guidance, and consultation can be provided remotely.
The authors of \cite{portnoy2020telemedicine} have enlisted the telemedicine providers, regulations, and have discussed the telehealth approach, its benefits, challenges, and the parameters that the clinicians need to look for to confirm the COVID-19 indicators. The major barriers is being the ignorance towards this mode and its usage, trustworthiness of traditional face-to-face communication and preference of interacting with known healthcare providers. The audio-visual assessment that the healthcare providers need to perform includes checking the temperature, ill appearance, calculation of the respiratory rate, presence or absence of cough, and other clinical symptoms. These enlisted parameters can be the focus of the Artificial Intelligence community to bring automation into this space.
A bot named ``Aapka Chikitsak'' \cite{bharti2020medbot} was designed using Natural Language Processing and voice-over techniques to cater the needs of the rural Indian population in this time of COVID-19 crisis. It is designed and developed with an intent to provide generic healthcare information, preventive measures, home remedies, and consultation for India-specific context with multi-lingual support to provide healthcare and wellbeing at no extra cost.
In addition to the physical diagnostic benefits, \cite{leite2020new} talks about the mitigation of psychological problems where psychotherapy sessions can be conducted using video calls.
The need for such a telehealth system is highlighted in \cite{khaleghi2020new},  as well as stating the primary advantages such as care for healthcare providers, avoiding overcrowding, and elderly care. However, this also raises the need for proper infrastructure and sufficient bandwidth to enable data transmission (audio, video, and images).

While the readiness of available voice assistants such as Alexa and Siri is analysed in \cite{sezgin2020readiness}, it is found that they need to be context aware and should be updated with the latest, reliable, and relevant content to be used as a part of a telehealth system for COVID-19. Also, the use of voice as a digital bio marker is missing.

\subsection{Chest X-Ray}
Outside of audio, speech, and language processing, imaging is a dominant signal processing and AI-based diagnosis method. 
The study performed in \cite{ai2020correlation} on 1\,014 cases found that the diagnosis using chest CT has a sensitivity of 97\,\% in the diagnosis of COVID-19.
Multiple efforts by several groups are put in the direction of developing a classifier to detect COVID-19 symptoms using chest X-Ray, such as those in \cite{toraman2020convolutional}, \cite{afshar2020covid}, \cite{apostolopoulos2020covid}, \cite{barstugan2020coronavirus}, \cite{elaziz2020new}, and many more. However, as stated in \cite{rubin2020role}, a multinational consensus is that imaging is indicated only for patients with worsening respiratory status. Hence, it is advised for only those patients who have moderate-severe clinical features and a high pre-test probability of disease. Hence, unlike in the Sections \ref{subsec:cough}, \ref{breathing}, and \ref{chatbot-S}, such measures are not suggested for early identification, and are preferred for diagnosis in a clinical setup. 

As concluded by \cite{wong2020frequency}, the chest X-ray findings in COVID-19 patients were found to be peaked at 10-12 days from symptom onset. Also, it is still required to visit a well-equipped clinical facility for such approaches. On a positive note, in the presence of mobile X-Ray machines, this approach can help in speeding up the diagnosis. The authors of \cite{li2020coronavirus} have found from an experimental outcome that the chest X-Ray may be useful as a standard method for the rapid diagnosis of COVID-19 to optimise the management of patients. However, CT is still limited for identifying specific viruses and distinguishing between viruses.
The smartphone application framework mentioned in \cite{maghdid2020novel} uses multiple sensor data such as that from a camera, microphone, fingerprint sensor, and inertial sensor for feeding CT scan images, human video tracking, as well as capturing cough sounds, temperature, and 30 seconds sit-to-stand movements to the lower layers for processing. With this, the authors plan to detect COVID-19 symptoms such as abnormalities in the CT-Scan images, nausea, fatigue, cough, and fever levels. They further use machine learning techniques such as a CNN and a RNN to derive the COVID-19 result based on the symptoms identified.

\section{Monitoring}
\label{sec:monitoring}

In an endeavour to detect the symptoms in each individual before a case gets serious, the government authorities in multiple countries had started examining and asking every individual if they have any COVID-19 symptoms. Such initiatives need a lot of human efforts to be invested. An alternative automation to accomplish such surveys can be using a system as described in \cite{lee2020carecall}. Using speech recognition, synthesis, and natural language understanding techniques, the CareCall system monitors individuals of Korea and Japan who had a contact with COVID-19 patients. This monitoring is done over a phone call using with and without human-in-the-loop process for three months. The system is used with over 13\,904 calls in which the authors reported 0\,\% false negative (self-reported COVID-19 subjects are not identified by the CareCall system) and 0.92\,\% false positive rate.

On the other hand, at an individual level, the precautions for stopping the virus spread include social distancing, wearing a mask, and maintaining respiratory hygiene. Especially at public places, the concerned authorities can monitor the population to confirm that the precautionary measures are adopted by them. This section describes such monitoring tools developed for the COVID-19 scenario.

\subsection{Face Mask Detection from Voice}

As described in Section \ref{subsec:cough}, collecting cough samples without covering mouth can lead to further spreading of the disease; mask wearing has to be mandated for donating a cough sample which requires an app to detect if the donor is wearing a mask or not. This year's Interspeech 2020 Computational Paralinguistics challengE (ComParE) \cite{schuller2020interspeech} featured a mask detection sub-challenge, where the task is to recognise whether the speaker was recorded while wearing a facial mask or not. The results from this challenge and the participants will be useful to develop a voice monitoring tool which detects the mask or no-mask of a speaker. Also, these algorithms serve as a pre-step for other speech processing algorithms such as breathing recognition, cough detection, or speech-based COVID-19 screening. 
One associated work in this direction is discussed in \cite{koike2020learning}, in which the authors have used techniques such as SpecAugment and mixup to generalise the deep models and have crossed the baseline by more than 4\,\% Unweighted Average Recall (UAR).
The winners of this sub-challenge \cite{szep2020paralinguistic} have used a deep convolutional neural network-based image classifier on the linear-scale 3-channel spectrograms of the speech segments to identify the mask wearing speakers. The authors have achieved a UAR of 80.1\,\% which is 8.3\,\% higher than the baseline using an ensemble of VGGNet, ResNet, and DenseNet architectures.

\subsection{Face Mask Detection from Image}
One of the precaution 
measures 
while stepping outside home suggested by the WHO to reduce the chance of getting infected or spreading COVID-19 is to wear a facial mask covering nose and mouth. 
Also, the mathematical analysis presented in \cite{eikenberry2020mask} suggests that wearing a mask can reduce community transmission and can help reduction in the peak death rate. Hence, similar to what has been outlined above, it is important that the concerned authorities keep a check on individuals wearing masks or not at public places, especially where the population is quite dense such as at airports, railway stations, and hospitals. 

The authors of \cite{wang2020masked} have provided the Masked Face Detection Dataset (MFDD), Real-world Masked Face Recognition Dataset (RMFRD), and the Simulated Masked Face Recognition Dataset (SMFRD) for the detection of masked and unmasked faces using image processing techniques. They have achieved 95\,\% accuracy in recognising masked faces. An Apple store app developed by LeewayHertz can be integrated with the existing setup of an 
Internet Protocol (IP) camera for getting alerts on detecting no-mask on a face. The system provides a facility to add phone numbers for receiving the alerts and also mechanism to see the face not wearing a mask for the admin.

\subsection{Cough and Breathing Analysis}

While at quarantine, the doctors need to monitor the cough history of patients, which can be done with a continuous cough monitoring device.
After we cross the crisis and organisations think of resuming the operations, continuous monitoring of common spaces such as canteens and lobbies can be realised to record COVID-19 specific coughs.
One such monitoring application is developed by the FluSense platform in \cite{al2020flusense}. It is a surveillance tool to detect influenza like illness from hospital waiting areas using cough sounds. 
Continuous monitoring and identification of abnormalities from the breathing rate has been done by \cite{wang2020abnormal} using image processing. A real-time application of cough detection is also applied using camera devices, which track and record the person who coughed, their location, and the number of coughs using a deep learning model for cough sound classification \footnote{https://healthcare-in-europe.com/en/news/covid-19-cough-camera-device-detects-location-of-coughing-sounds-in-real-time.html}. The authors achieved a test accuracy of 87.4\,\% in identifying cough sounds in an office environment. However, they have not tried identifying COVID-19 cough sounds.

\subsection{Mental Health -- Emotion Detection}

The disease spread has equally affected the physical and mental health of the individuals, which is primarily due to plenty of mandatory precautionary measures such as social-distancing, work from home and the quarantine procedures which usually takes around 15 days for the patients to be alone. Also, the health care providers owing to their hectic routines followed by quarantine days are subject to undergo mental health issues. To cater for the growing need of addressing this issue, not only is there a high demand, but the physical presence of the available psychologists is also missing. As found in \cite{patel2020rapid}, the COVID-19 pandemic has generated unprecedented demand for tele-health based clinical services. 

Among several initiatives taken against mental health issues such as stress, anxiety, and depression, we are yet to see these emotions being analysed from the speech signal during the COVID-19 period. This demands for the relevant data. Very recently, a study is conducted by the authors of \cite{han2020early} on the speech signal of COVID-19 diagnosed patients. The 
behavioural 
parameters detected from speech includes, sleep quality, fatigue, and anxiety considering corresponding self-reported measures as ground truth and have achieved an average accuracy of 0.69 
in estimating the severity of COVID-19.
The authors have collected 5 speech recordings from each of the 52 patients while speaking neutral statements along with 3 responses to the questions regarding their sleep quality, fatigue and anxiety. The metadata of these patients consist of age, gender, height, and weight. The authors report UAR of 0.61\,\%, 0.46\,\% and 0.56\,\% for the three tasks sleep quality, fatigue and anxiety respectively. They have extracted eGeMAPS feature set using OpenSMILE toolkit \cite{eyben2010opensmile} and used SVM classifier .

This year's ComParE challenge at Interspeech 2020 \cite{schuller2020interspeech} had an Elderly Emotion Detection sub-challenge, where the speech captured from elderly 
subjects had to be classified into low, medium, and high valence and arousal. 
It was found that specific age groups such as that of
infants \cite{li2020mini} and
elderly above 60 years are more prone to the infection, due to which this age group presumably largely needs to follow the restrictions posed by the pandemic for a larger period in the future as well. This shows that it will be crucial to understand the effects of this phase on their psychological parameters such as emotions. 
The winner of this challenge, \cite{sogancioglu2020everything} used acoustic features for arousal classification and linguistic features for valence classification. The authors used fisher vector encoding of the 76 dimensional acoustic features comprising of 24 MFCCs and the rest of the Perceptual Linear Prediction cepstrum (RASTA-PLP) for 12th order linear prediction, together with their temporal delta coefficients. For arousal detection, the ensemble of these features combined with the baseline features gave UAR of 57.5\,\%, which is 7.1\,\% higher than the baseline.
For valence detection, the authors used an ensemble of TF-IDF features, FastText+Polarity features and Dictionary-based features to achieve the UAR of 62.3\,\%.

\subsection{Text Analysis}
Text data from several individuals on social media, hospitality feedback from the patients on the treatment given to them, and forums such as blogs provide useful information about different COVID-19 related aspects. The mental health support communities on Reddit are analysed in \cite{biester2020quantifying}, to understand the effects of COVID-19 on the mental health of society. It is observed that the posts on this community forum per day have increased specifically in the subreddits of Anxiety and SuicideWatch, however, decrease in the subreddit of depression during the pandemic period. Similarly, in \cite{ahmed2020dangerous}, the twitter data with 5GCoronavirus hashtag is analysed to understand the drivers of the 5G COVID-19 conspiracy theory and strategies to help in reducing the spread of mis-information circulating in society.

The authors of \cite{guney2020using} have used machine learning techniques to categorise sentiments from the responses to Press Ganey patient experience surveys. From this analysis, it is found that the patients have expressed negative comments about the cleanliness and logistics and have given positive comments about their interactions with clinicians. 

A text-mining based study on the impact of COVID-19 on the individuals of Paris and France from April 23 to June 18, 2020 is studied by the authors of \cite{saire2020study}. The authors used twitter data, having 1,496,375 tweets from 285,114 users, specific to the given dates, location, French language and COVID-19. They found a decreasing pattern of publications/interest, and an increased impact on the health crisis and economy generated by coronavirus. Another study done by the authors of \cite{koh2020loneliness}, on twitter data to understand the impact of COVID-19 on loneliness among individuals. The authors found that the highly influential users were talking more about the mental health effects of loneliness during COVID-19.
To accelerate the text-mining based research on the COVID-19 related publications, \cite{wang2020cord} CORD-19 is the dataset formed of research paper from PubMed Central, bioRxiv and medRxiv preprint servers, WHO Covid-19 Database, Elsevier and Springer Nature under special Covid-19 open access licenses. 92\,\% of the papers are from Biology, Medicine, and Chemistry domains. 

\section{Awareness}
\label{sec:awareness}
Speech recognition and synthesis algorithms have been widely used in the development of chat-bots to provide human like interactions. In this time of crisis, chat-bots are helping in spreading valuable information about COVID-19 to end users. Once an infected gets cured, they can help researchers with not only their experience but also with donating components such as plasma. CDC has been encouraging recovered individuals to donate their plasma for development of blood related therapies. For collection of plasma, Microsoft has developed the chat-bot \footnote{https://covig-19plasmaalliance.org} which interacts with individuals to gather the required information such as, the duration for which they are tested negative, their age, weight, and also takes their pin-code to help them know their nearest donation center. 



\section{Next Steps and Challenges}
\label{sec:nextsteps}
With the smartphone likely being the most convenient and available asset that every individual carries all the time, more of smartphone-based applications for detecting COVID-19 symptoms will help in controlling the virus spread. The authors of \cite{albes2020squeeze} have addressed the memory and power consumption issues for importing a deep learning model for detection of cold from the speech signal. They propose network pruning and quantisation techniques to reduce the model size, with which they achieved a size reduction of 95\,\% in MBytes without affecting the recognition performance.

Along with the physical COVID-19 symptoms, the behavioural aspects also need greater attention.
Since the work culture is moving more towards working-from-home, it will be required to detect some behavioural parameters such as fatigue and sleepiness for the self monitoring of the working professionals. 
The authors of \cite{schuller2019interspeech} had organised a challenge at Interspeech 2019, for detecting the sleepiness among individuals where a labelled dataset of 950 subjects with Karolinska Sleepiness Scale score is provided. The winners of this challenge \cite{gosztolya2019using} had reported the spearman's correlation coefficient of 0.383 on the test dataset using COMPARE features, Bag of Audio Words (BoAW), and Fisher vectors. The same dataset is used in \cite{amiriparian2020novel}, in which the feature representation done by attention and autoencoder techniques are fused to train a support vector regressor. The authors report Spearman’s correlation coefficients of 0.367 on the test dataset.

The behavioural parameters such as stress, anxiety and depression needs greater attention to be paid in these days. 
An audio-visual emotion detection challenge organised by \cite{ringeval2017avec} provides a speech dataset named, ''Distress Analysis Interview Corpus – Wizard of Oz (DAICWOZ)'' which is a part of a larger corpus, the ''Distress Analysis Interview Corpus'' (DAIC) \cite{gratch2014distress}, that contains clinical interviews designed to support the diagnosis of psychological distress conditions such as anxiety, depression, and post-traumatic stress disorder. In DAICWOZ, the participants rate themselves on the eight-item Patient Health Questionnaire depression scale (PHQ-8).
The authors of \cite{zhao2020hierarchical}, have worked with the same dataset using Hierarchical Attention Transfer Network and report an RMSE of 5.66 and an Mean Absolute Error of 4.28 on the test dataset.

A major challenge given the social distancing norms is getting the relevant and accurate speech data for developing machine learning models. Speech enabled chat-bots can play a significant role in this. There are other challenges as well from the design perspective of chat-bots. The authors of \cite{miner2020chatbots} have expressed both positive effects and drawbacks associated with using chatbots for sharing the latest information, encouraging desired health impacting 
behaviours, 
and reducing the psychological damage caused by fear and isolation. This shows that the design of chat-bots should be well thought of for using them, otherwise, they might have negative impact as well. 
An optimistic approach in these difficult times has been to work towards safe and secure environment for the post pandemic situation so that the society gains the trust and confidence back. This shows the need of accurate and reliable screening and monitoring measures at public places. 

\section{Conclusion}
\label{sec:conclusion}

Speech and human audio analysis is found to be predominantly useful for COVID-19 analysis.
Several initiatives towards identifying cough sound and distinguishing COVID-19 cough from other illnesses are currently taken. It looks promising that soon a cough sound-based sufficiently accurate COVID-19 detector for several real-world use-cases will be available. Such detectors, when integrated with chat-bots can enhance the screening, diagnosing, and monitoring efforts with reduction in human interventions.
Further research is required for breathing and speech signal-based COVID-19 analysis, where it is more important to identify the exact bio-markers. With increasing correlation established between speech and breathing signals, detecting breathing disorders from the speech signals will be useful.
In many countries, a second, and even a third wave of COVID-19 infection has been found to occur infecting many more individuals. This suggests for the urgent need of robust monitoring mechanisms. 
Many elderly individuals have been inside home for almost the entire year.
The past research on the detection of OSA and stress needs to be taken forward in the COVID-19 context for the elderly population.
Besides, promising applications for the usage of language processing and other signal analyses have been shown. In sum, we are positive that the combination of intelligent audio, speech, language, and other signal analysis can help make an important contribution in the fight against the COVID-19 and oncoming similar pandemics -- alone, or in combination with other methods. 


\section{Acknowledgements}
We would like to thank all researchers, health supporters, and others helping in this crisis. Our hearts are with those affected and their families and friends.  
We acknowledge funding from the German BMWi by ZIM grant No.\ 16KN069402 (KIrun).

%
%

\bibliographystyle{elsarticle-num}
\bibliography{mybib}   



\end{document}